\documentstyle[12pt,epsfig,graphics,cite,amssymb,amsmath,bm,cool,caption,booktabs]{article}
\newcommand{\ups}{\upsilon}
\begin{document}\bibliographystyle{plain}\begin{titlepage}
\renewcommand{\thefootnote}{\fnsymbol{footnote}}\hfill
\begin{tabular}{l}HEPHY-PUB 1012/18\\December 2018\end{tabular}\\
[2cm]\Large\begin{center}{\bf THE SPINLESS RELATIVISTIC HELLMANN
PROBLEM}\\[1cm]\large{\bf Wolfgang LUCHA\footnote[1]{\normalsize\
\emph{E-mail address\/}: Wolfgang.Lucha@oeaw.ac.at}}\\[.3cm]
\normalsize Institute for High Energy Physics,\\Austrian Academy
of Sciences,\\Nikolsdorfergasse 18, A-1050 Vienna, Austria\\[1cm]
\large{\bf Franz F.~SCH\"OBERL\footnote[2]{\normalsize\
\emph{E-mail address\/}: franz.schoeberl@univie.ac.at}}\\[.3cm]
\normalsize Faculty of Physics, University of Vienna,\\
Boltzmanngasse 5, A-1090 Vienna, Austria\\[2cm]{\normalsize\bf
Abstract}\end{center}\normalsize We compile some easily deducible
information on the discrete eigenvalue spectra of spinless
Salpeter equations encompassing, besides a relativistic kinetic
term, interactions which are expressible as superpositions of an
attractive Coulomb potential and an either attractive or repulsive
Yukawa potential and, hence, generalizations of the Hellmann
potential employed in several areas of science. These insights
should provide useful guidelines to all attempts of finding
appropriate descriptions of bound states by (semi-) relativistic
equations of~motion.

\vspace{3ex}

\vspace{2ex}


\renewcommand{\thefootnote}{\arabic{footnote}}\end{titlepage}

\section{Spinless-Salpeter-Based Approach to Bound States}Within
quantum field theory, the \emph{homogeneous Bethe--Salpeter
equation\/} \cite{BSEa,BSEb,BSEc} constitutes a
Poincar\'e-covariant approach to bound states. Driven by the
desire to describe bound states to the utmost reasonable extent by
analytic tools, a variety of directions has been proposed for the
diminution of the complexity inherent to the Bethe--Salpeter
formalism. Performing a three-dimensional reduction, by assuming
for the involved bound-state constituents both propagation like
free particles and instantaneity of their mutual interactions, and
dropping any negative-energy contribution and all bound-state
constituents' spin degrees of freedom takes us to the spinless
Salpeter equation: a semirelativistic bound-state equation that
may be formulated as the eigenvalue equation of an appropriate
Hamiltonian $H$ composed of its relativistic kinetic energy $T$
and some interaction potential $V$. For bound states of only two
constituents of relative momentum $\bm{p}$, relative coordinate
$\bm{x}$, and masses $m$, for simplicity of notation here chosen
to be equal, this operator $H$ is (in natural
units~$\hbar=c=1$)~of~the~form\begin{equation}H\equiv
T(\bm{p})+V(\bm{x})\ ,\qquad T(\bm{p})\equiv2\,\sqrt{\bm{p}^2+m^2}
\ .\label{H}\end{equation}This class of Hamiltonians may be viewed
as immediate generalization of its nonrelativistic Schr\"odinger
counterpart towards incorporation of relativistic kinematics,
albeit deduced~at the price of having to cope with the nonlocality
induced by the ``relativized'' kinetic energy.

The nonlocality of any semirelativistic Hamiltonian $H$ such as
the one of Eq.~(\ref{H}) poses a grave obstacle to the preferred
analytic solvability of the corresponding equation of motion.
Consequently, when attempting to pin down the solutions to such
bound-state equation by whatever approach seems best to one,
reliable guidelines, in form of rigorous constraints on any
bound-state spectrum to be expected, should be highly welcome.
Needless to~say,~upon having a set of solutions at one's disposal,
its significance can be easily examined by various really powerful
tools, for instance, the generalization of the virial theorem of
nonrelativistic quantum theory to the incorporation of
relativistic kinematics \cite{WL:RVT,WL:RVTs}. Utilizing
techniques of this kind has proven extremely efficient in the
separation of the wheat from the chaff
\cite{WL:WS,WL:Q@W,WL:H,WL:Y,WL:K}.

In the present analysis, we adopt various considerations of the
above type (to the~extent applicable) for a characterization of
what can be dubbed the \emph{relativistic Hellmann problem},
obtained by specifying the interaction term $V(\bm{x})$ in our
Hamiltonian (\ref{H}) to be a spherically symmetric potential
$V(r)$ ($r\equiv|\bm{x}|$) of a form that generalizes the shape
originally proposed by Hellmann \cite{H,H+} and that has found
widespread application in different areas of physics and
chemistry. By definition, that generalization of Hellmann's
potential is a superposition\begin{align}V(\bm{x})=V_{\rm
H}(r)&\equiv V_{\rm C}(r)+V_{\rm Y}(r)\nonumber\\
&=-\frac{\kappa}{r}-\ups\,\frac{\exp(-b\,r)}{r}\ ,\qquad\kappa\ge0
\ ,\qquad\ups\gtreqqless0\ ,\qquad b>0\ ,\label{VH}\end{align}of
a, by convention, attractive Coulomb term $V_{\rm C}(r)$ with
related coupling parameter~$\kappa\ge0$,$$V_{\rm C}(r)=-
\frac{\kappa}{r}\ ,\qquad\kappa\ge0\ ,$$and a Yukawa term $V_{\rm
Y}(r)$ with related coupling strength $\ups\gtreqqless0$ and range
parameter~$b>0$,$$V_{\rm Y}(r)=-\ups\,\frac{\exp(-b\,r)}{r}\
,\qquad\ups\gtreqqless0\ ,\qquad b>0\ .$$In this spirit, after
listing all possible shapes of $V_{\rm H}(r)$ (Sec.~\ref{C}), we
explore the circumstances under which $H$ is bounded from below
(Sec.~\ref{BB}), evaluate our chances of limiting the number of
bound states (Sec.~\ref{N}), constrain their energies
(Sec.~\ref{U}), and reassess our findings (Sec.~\ref{I}).

\section{Classification: Possible Hellmann-Potential Shapes}
\label{C}
\begin{figure}[ht]\begin{center}\begin{tabular}{ccc}
\psfig{figure=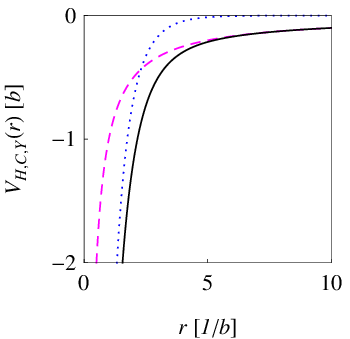,scale=1.40799}&
\psfig{figure=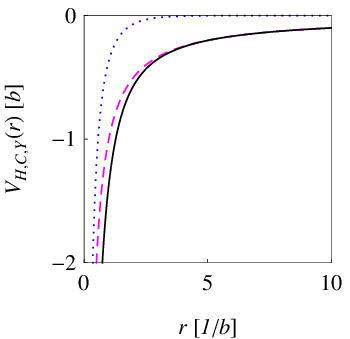,scale=1.40799}&
\psfig{figure=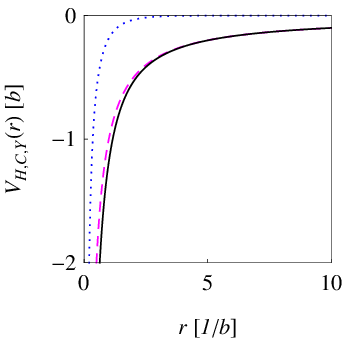,scale=1.40799}\\[1ex](a)&(b)&(c)\\[2ex]
\psfig{figure=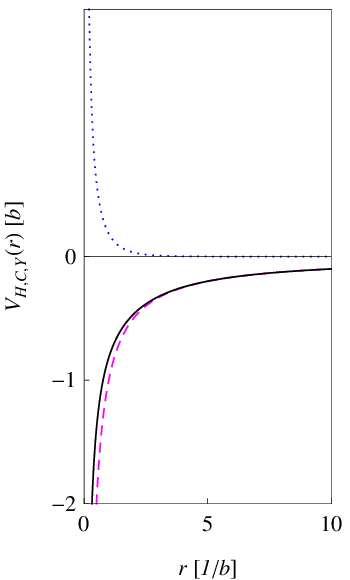,scale=1.40799}&
\psfig{figure=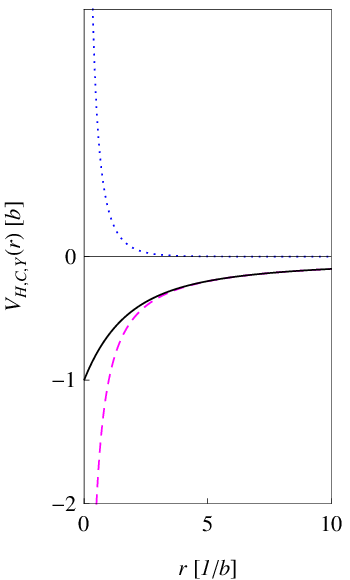,scale=1.40799}&
\psfig{figure=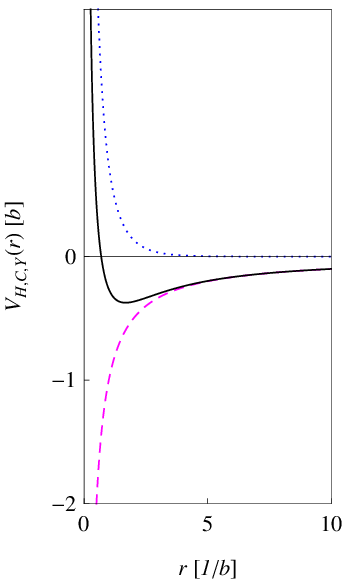,scale=1.40799}\\[1ex](d)&(e)&(f)
\end{tabular}\caption{Hellmann potential $V_{\rm H}(r)$ (black
solid lines), defined by the superposition (\ref{VH}) of a (by
assumption attractive) Coulomb potential $V_{\rm C}(r)$ (dashed
magenta lines) and a Yukawa potential $V_{\rm Y}(r)$ (dotted blue
lines), for a fixed value $\kappa=1$ of the Coulomb~coupling,~and
six choices, $\ups=10$ (a), $\ups=1$ (b), $\ups=0.5$ (c),
$\ups=-0.5$ (d), $\ups=-1$ (e), and $\ups=-2$ (f), of the Yukawa
coupling, in one-to-one correspondence representative of the size
relationships~(\ref{rel}).}\label{V}\end{center}\end{figure}

Let us start by examining the gross behaviour of our
potential $V_{\rm H}(r)$ for $r\to\infty$ and~$r\to0$.
\begin{itemize}\item In the limit $r\to\infty$, any Yukawa
potential $V_{\rm Y}(r)$ vanishes, due to its rapidly decaying
exponential factor, faster than the Coulomb potential $V_{\rm
C}(r)$. Accordingly, the large-$r$ behaviour of the Hellmann
potential $V_{\rm H}(r)$ inevitably approaches that of its Coulomb
component $V_{\rm C}(r)$, independently of the values of the
Yukawa parameters $\ups$ and $b>0$:$$V_{\rm
H}(r)\xrightarrow[r\to\infty]{}-\frac{\kappa}{r} \equiv V_{\rm
C}(r)\ .$$\item The $V_{\rm H}(r)$ behaviour near the origin is
determined by the size of $\ups$ relative to
$-\kappa\le0$:$$\lim_{r\to0}V_{\rm H}(r)=\left\{\begin{tabular}
{lcrcr}$-\infty$&&$\kappa+\ups>0$&&$\ups>-\kappa\ ,$\\$\ups\,b
=-\kappa\,b$&$\quad$for&$\kappa+\ups=0$& $\Longleftrightarrow$&
$\ups=-\kappa\ ,$\\$+\infty$&& $\kappa+\ups<0$&&$\ups<-\kappa\ .$
\end{tabular}\right.$$\end{itemize}There exist some further
peculiar features of the Hellmann potential (\ref{VH}) worth
mentioning:\begin{itemize}\item The case of sufficiently negative
Yukawa coupling, more exactly, the interval $\ups<-\kappa$,
defines the only instance where $V_{\rm H}(r)$ develops a
repulsive core, enabled by its zero~at$$r_0=-\frac{1}{b}
\ln\!\left(-\frac{\kappa}{\ups}\right)=\frac{1}{b}
\ln\!\left(-\frac{\ups}{\kappa}\right),$$and, accordingly, a
global minimum $V_{\rm H}(\underline{r})$, situated at the
straightforward solution~of$$\ups\,(1+b\,\underline{r})
\exp(-b\,\underline{r}) =-\kappa\ .$$\item For sufficiently
positive Yukawa coupling, that is to say, for $\ups>\kappa$, we
find a crossover of the Coulomb and Yukawa contributions to the
Hellmann potential for finite $r$,~\emph{viz.},$$r_\times=-
\frac{1}{b}\ln\frac{\kappa}{\ups}=\frac{1}{b}\ln\frac{\ups}{\kappa}
\ .$$\end{itemize}In more detail, we may classify the behaviour of
the Hellmann potential into six intervals of the Yukawa coupling
$\ups$ (leaving aside the value $\ups=0$ reducing $V_{\rm H}(r)$
to the Coulomb~case):\begin{subequations}\begin{align}
\ups>\kappa&\qquad\Longrightarrow\qquad\kappa+\ups>0\
,\label{ygk}\\
\ups=\kappa&\qquad\Longrightarrow\qquad\kappa+\ups>0\
,\label{y=k}\\
0<\ups<\kappa&\qquad\Longrightarrow\qquad\kappa+\ups>0\
,\label{ylk}\\
-\kappa<\ups<0&\qquad\Longrightarrow\qquad\kappa+\ups>0\
,\label{yg-k}\\
\ups=-\kappa&\qquad\Longrightarrow\qquad\kappa+\ups=0\
,\label{ye-k}\\
\ups<-\kappa&\qquad\Longrightarrow\qquad\kappa+\ups<0\
.\label{yl-k}\end{align}\label{rel}\end{subequations}Figure
\ref{V} shows an illustrative example for each distinct potential
shape, with the plots~\ref{V}(a) through \ref{V}(f) corresponding
bijectively to the $\ups$ regions (\ref{ygk}) through
(\ref{yl-k}). For a vanishing Yukawa coupling, the Hellmann
potential reduces, trivially, to the Coulomb potential. The
spinless relativistic problem posed by the latter has been
analyzed very extensively \cite{IWH,IWH+,MR,RRSMS,WL:C}.

\section{Boundedness from Below: Lower Operator Bounds}\label{BB}
Before being entitled to talk about bound states controlled by a
Hamiltonian $H$, one should feel obliged to establish the
appropriate semiboundedness of the operator under discussion,
defined, for given kinetic term $T(\bm{p})$, by the behaviour of
the involved interaction potential, in our case of the Hellmann
potential $V_{\rm H}(r)$. Unsurprisingly, the latter potential is
bounded from below by a Coulomb-type term with suitably chosen
coupling strength. Depending on whether the Yukawa contribution to
$V_{\rm H}(r)$ is positive or negative, we distinguish two~cases:
\pagebreak
\begin{alignat*}{2}&V_{\rm H}(r)\ge-\frac{\kappa+\ups}{r}\ ,&\qquad
\mbox{for}\qquad\ups>0\qquad\Longleftrightarrow\qquad V_{\rm
Y}(r)<0\ ,\\[1ex]&V_{\rm H}(r)\ge-\frac{\kappa}{r}\ ,&\qquad
\mbox{for}\qquad\ups\le0\qquad\Longleftrightarrow\qquad V_{\rm
Y}(r)\ge0\ .\end{alignat*}For Yukawa couplings in the range $\ups
\le-\kappa$ entailing the positivity of the Yukawa potential, this
Coulomb lower bound is clearly far from optimum and consequently
easily improvable:\begin{itemize}\item For $\ups<-\kappa$, the
Hellmann potential is trivially bounded from below by its minimum,
\begin{equation}V_{\rm H}(r)\ge\min_{0\le r<\infty}V_{\rm H}(r)=
V_{\rm H}(\underline{r})>-\infty\ .\label{Sf}\end{equation}\item
For $\ups=-\kappa$, the Hellmann potential is bounded from below
by its value at the~origin,\begin{equation}V_{\rm H}(r)\ge V_{\rm
H}(0)=\ups\,b =-\kappa\,b\ .\label{Se}\end{equation}\end{itemize}
For $\ups>-\kappa$, we may take advantage of available knowledge
about the spectral properties of the semirelativistic Hamiltonian
$H$ for Coulomb-type interaction potentials $V_{\rm C}(r)$ with
yet to be constrained strength $\kappa$; the eigenvalue equation
of this operator $H$ defines the spinless relativistic Coulomb
problem. The relevant quintessence of spectral insights
\cite{IWH,IWH+} is that,\begin{itemize}\item for $V(\bm{x})=V_{\rm
C}(r)$, the Hamiltonian $H$ is essentially self-adjoint for
Coulomb coupling$$\kappa\le1\ ;$$\item if $V(\bm{x})=V_{\rm
C}(r)$, the Hamiltonian $H$ has a Friedrichs extension for
Coulomb~coupling$$\kappa\le\kappa_{\rm c}=\frac{4} {\pi}=
1.273239\dots\ ;$$\item for $V(\bm{x})=V_{\rm C}(r)$, the spectrum
of the Hamiltonian $H$, $\sigma(H),$ is bounded from~below:
$$\sigma(H)\ge2\,m\, \sqrt{1-\left(\frac{\kappa}{\kappa_{\rm
c}}\right)^{\!2}}=2\,m\,\sqrt{1-\left(\frac{\pi\,\kappa}{4}
\right)^{\!2}}\qquad\mbox{if and only if}\qquad\kappa\le
\kappa_{\rm c}\ .$$\end{itemize} Restricting the admissible
interval of couplings $\kappa$ yields an improvement of
this~bound~\cite{MR}:\begin{equation}\sigma(H)\ge2\,m\,
\sqrt{\frac{1+\sqrt{1-\kappa^2}}{2}}\qquad\mbox{for}\qquad\kappa
\le1\ .\label{SM}\end{equation}So, in conclusion, we managed to
deduce a first firm intermediate result: By limiting, where
necessary, the magnitude of both coupling parameters involved in
the interaction potential, the spinless relativistic Hellmann
problem under study may be formulated in a well-defined manner.
Phrased differently, it can be defined in such a way that its
spectrum~is~guaranteed to be bounded from below, irrespective of
the range parameter entering in its Yukawa term.

By evaluating expectation values of the Hamiltonian $H$ over
cleverly chosen trial states, it is a rather easy task both to
convince oneself of the necessity of constraining those among the
involved interaction strengths that --- by contributing to a
negative singularity of $V_{\rm H}(r)$ at the origin --- otherwise
will spoil the desired boundedness from below of the Hamiltonian
$H$ and to provide (more or less satisfactory) estimates for upper
bounds to those couplings. For all potentially problematic Yukawa
coupling strengths $\ups>-\kappa$, this is demonstrated by
adopting a rather simple trial state defined, in either
configuration~($r\equiv|\bm{x}|$) or momentum ($p\equiv|\bm{p}|$)
space, by the normalized radial function (enjoying a variational
parameter $\mu>0$)\pagebreak
$$\phi(r)=2\,\mu^{3/2}\exp(-\mu\,r)\qquad\Longleftrightarrow\qquad
\widetilde\phi(p)=\frac{(2\,\mu)^{5/2}}{\sqrt{\pi}}\,\frac{1}
{(p^2+\mu^2)^2}\ .$$The implied expectation value of the kinetic
term $T(\bm{p})$ may be given once and forever \cite{WL:H,WL:Y}:
\begin{align*}\langle T(\bm{p})\rangle&=
\frac{4}{3\,\pi\,(m^2-\mu^2)^{5/2}}\\&\times
\left[\mu\,\sqrt{m^2-\mu^2}\,(3\,m^4-4\,m^2\,\mu^2+4\,\mu^4)+3\,m^4
\,(m^2-2\,\mu^2)\,\ArcSec{\frac{m}{\mu}}\right].\end{align*}Likewise,
that of any sufficiently simple potential, such as $V_{\rm H}(r)$,
can be found analytically:$$\langle V_{\rm H}(r)\rangle=
-\kappa\,\mu-\frac{4\,\ups\,\mu^3}{(b+2\,\mu)^2}\ .$$By Laurent
expansion in $1/\mu$, the emerging expectation value of $H$
behaves, for large~$\mu$,~like\begin{align*}\langle H\rangle&
\equiv\langle T(\bm{p})\rangle+\langle V_{\rm H}(r)\rangle\\[1ex]&
=\left(\frac{16}{3\,\pi}-\kappa-\ups\right)\mu+\ups\,b+O\!\left(
\frac{1}{\mu}\right)\xrightarrow[\mu\to\infty]{}-\infty\qquad
\mbox{for}\qquad\kappa+\ups>\frac{16}{3\,\pi}\ .\end{align*}
Mitigating the Coulombic singularity of $V_{\rm H}(r)$ at the
origin needed for a finite ground-state energy thus requires the
sum of Coulomb and Yukawa couplings to satisfy an upper~bound,
$$\kappa+\ups\le\frac{16}{3\,\pi}=1.69765\dots\ ,$$that can be
sharpened by discriminating with respect to the sign of our Yukawa
coupling~$\ups$:\begin{alignat*}{2}&\kappa+\ups\le\frac{16}{3\,\pi}
&&\qquad\mbox{for}\qquad\ups=|\ups|>0\ ,\\[1ex]&\kappa\le\frac{16}
{3\,\pi}+|\ups|&&\qquad\mbox{for}\qquad-\kappa<\ups=-|\ups|<0\
.\end{alignat*}

\section{Delimiting Spinless-Salpeter Bound-State Counts?}\label{N}
Another set of questions entering one's mind concerns the actual
\emph{number\/} of bound states of spinless Salpeter equations. Is
this number nonvanishing? Does there exist a bound~state at all?
Is this number finite or infinite? A couple of considerations may
shed light on~this~issue:\begin{itemize}\item For the Hamiltonian
(\ref{H}) with relativistic kinetic term $T(\bm{p})$, an
overwhelming amount of knowledge about the number of its bound
states does not seem to exist. The~bound found in Ref.~\cite{ICD}
holds for every nonpositive interaction potential $V(\bm{x})$ that
satisfies$$V(\bm{x})\in L^{3/2}({\mathbb R}^3)\cap L^3({\mathbb
R}^3)\ .$$Consequently, this latter finding is straightforwardly
applicable to the Woods--Saxon potential \cite{WL:WS} or the
kink-like potential \cite{WL:K} but not to the Hellmann potential
(\ref{VH})~since$$V_{\rm H}(r)\notin L^{3/2}({\mathbb R}^3)\cap
L^3({\mathbb R}^3)\ .$$\item For any potential $V(\bm{x})$, the
operator inequality fulfilled by the kinetic term~$T(\bm{p})$~and
its nonrelativistic (NR) limit $T_{\rm NR}(\bm{p})$ translates,
via the arising Hamiltonians,~into~an inequality for the
\emph{discrete\/} eigenvalues at the bottom of the associated
energy spectra: The relativistic kinetic energy $T(\bm{p})$ is
\emph{bounded from above\/} by its nonrelativistic~limit,
$$T(\bm{p})\le T_{\rm NR}(\bm{p})\equiv2\,m+\frac{\bm{p}^2}{m}\
.$$So, the spinless-Salpeter Hamiltonian $H$ and its Schr\"odinger
counterpart $H_{\rm NR}$ satisfy$$H\le H_{\rm NR}\equiv T_{\rm
NR}(\bm{p})+V(\bm{x})\ .$$A spectral comparison theorem, reviewed,
\emph{e.g.}, in Refs.~\cite{WL:RCP,WL:T1,WL:T2,WL99,WL:TWR}, then
allows~us~to~cast this operator inequality into a relation among
the corresponding ordered eigenvalues:$$E_k\le E_{k,{\rm NR}}\
,\qquad k\in{\mathbb N}_0\ .$$Hence, to every discrete bound state
of $H_{\rm NR}$ corresponds a discrete bound state of $H.$ The
number of bound states of $H$ thus \emph{cannot be less\/} than
that of the respective~$H_{\rm NR}.$\item Every such Schr\"odinger
Hamiltonian with interaction potential (\ref{VH}) of Hellmann
type, in turn, upon being confronted with suitable Coulomb
problems, proves to support at least one discrete bound state,
irrespective of the value of the Yukawa~coupling~$\ups$ \cite{HK}.
\item Owing to the comparatively fast decay of the exponential in
its Yukawa contribution, the behaviour of any Hellmann potential
approaches, in the limit of large distances $r$, the behaviour of
its Coulomb contribution. For a nonrelativistic
Hamiltonian~$H_{\rm NR}$, in turn, the approach of an attractive
Coulomb-like potential to zero in the limit~$r\to\infty$ is
definitely not fast enough to constrain the number of bound states
to a finite value. This circumstance should be and is reflected by
existing upper bounds on the number of nonrelativistic bound
states. For instance, the limit of Ref.~\cite{VB} diverges since,
for a Hellmann potential, the integral entering this finding as
the key quantity~behaves~like$$\int_0^R{\rm d}r\,r\,|\widetilde
V_{\rm H}(r)|\xrightarrow[R\to\infty]{}\infty\ ,\qquad\widetilde
V_{\rm H}(r)\equiv-\max[0,-V_{\rm H}(r)]\ .$$\end{itemize}In
summary, our tentative conclusion must be that (as consequence of
the predominance of the Coulomb potential at sufficiently large
distances) the comparatively slow Coulomb-like approach to zero of
any Hellmann potential causes --- in contrast to the Yukawa case
--- the number of bound states of a spinless relativistic Hellmann
problem to grow~beyond~bounds.

\section{Variational Upper Limits on Bound-State Energies}
\label{U}Given the lower bounds of Sec.~\ref{BB}, we next aim at
constricting, as tight as possible, the exact eigenvalues of the
semirelativistic Hamiltonian $H$ with generalized Hellmann
potential (\ref{VH}).

A main tool for a rough localization of the discrete spectrum of a
Hilbert-space operator $H$, consisting of all isolated eigenvalues
of finite multiplicity below the onset of the essential spectrum,
is the Rayleigh--Ritz variational technique provided by the
minimum--maximum theorem \cite{RS}: The lowest-lying $d$ among the
eigenvalues, $E_k$, $k\in{\mathbb N}_0$, assumed to be ordered
according to $E_0\le E_1\le\cdots$, of a \emph{self-adjoint\/}
operator $H$ bounded from below~are bounded from above by the $d$
eigenvalues, $\widehat E_k$, $k=0,1,\dots,d-1$, assumed to be
ordered according~to $\widehat E_0\le\widehat
E_1\le\cdots\le\widehat E_{d-1}$, of the restriction of $H$ to any
$d$-dimensional subspace of its domain,\pagebreak$$E_k\le\widehat
E_k\qquad \forall\quad k=0,1,\dots,d-1\ .$$

For a spherically symmetric interaction potential
$V(\bm{x})=V(r)$, each basis vector of the finite-dimensional
trial space setting the stage for application of this variational
framework may be conveniently represented by a product of a radial
function and a spherical harmonic ${\cal Y}_{\ell m}(\Omega)$ for
orbital angular momentum and associated projection quantum numbers
$\ell\in{\mathbb N}_0$ and $m=\ell,\ell-1,\dots,-\ell$, depending
on the solid angle $\Omega.$ An orthonormal basis of this kind,
rather well suited for tackling on analytical grounds bound-state
problems whose definition comprises the square-root operator of
the relativistic kinetic energy, utilizes two variational
parameters: $\mu$, with dimension of mass and satisfying
$\mu\in(0,\infty)$, and $\beta,$ dimensionless and satisfying
$\beta\in(-\frac{1}{2},\infty)$. It can be given analytically in
configuration and momentum space, in the former by
generalized-Laguerre orthogonal polynomials $L_k^{(\gamma)}(x)$
\cite{AS} for parameter~$\gamma$:\begin{align}\psi_{k,\ell
m}(\bm{x})&=\sqrt{\frac{(2\,\mu)^{2\ell+2\beta+1}\,k!}
{\Gamma(2\,\ell+2\,\beta+k+1)}}\,r^{\ell+\beta-1}\exp(-\mu\,r)\,
L_k^{(2\ell+2\beta)}(2\,\mu\,r)\,{\cal Y}_{\ell
m}(\Omega_{\bm{x}})\ ,\nonumber\\[1ex]&L_k^{(\gamma)}(x)\equiv
\sum_{t=0}^k\,\binom{k+\gamma}{k-t}\frac{(-x)^t}{t!}\ ,\qquad
k=0,1,2,\dots\ ,\nonumber\\[1ex]&\int{\rm d}^3x\,\psi_{k,\ell
m}^\ast(\bm{x)}\, \psi_{k',\ell'm'}(\bm{x)}=\delta_{kk'}\,
\delta_{\ell\ell'}\, \delta_{mm'}\ .\label{L}\end{align}Fourier
transformation immediately provides the momentum-space
representation of these basis states; explicit expressions of
these functions can be found in, \emph{e.g.},
Refs.~\cite{WL:T1,WL:TWR,WL:WS,WL:H,WL:Y}.

Table~\ref{T} illustrates the application of the variational
technique recalled above to spinless relativistic Hellmann
problems by presenting the set of upper limits on the
\emph{binding\/} energies\begin{equation}B_k\equiv E_k-2\,m\
,\qquad k=0,1,2,\dots,d-1\ ,\label{B}\end{equation} derived, for
representatives of three of our six categories of generalized
Hellmann potentials identified by Eqs.~(\ref{rel}), for a kind of
self-suggesting choice of parameter values and optimized
(\emph{i.e.}, reasonably chosen) trial-space dimension
$d\in\{29,30,31\}$. Enlarging the trial space~as well as
exploiting one's freedom to vary the two parameters $\mu$ and
$\beta$ clearly should allow~for an improvement of any such upper
limits. A dash in Table~\ref{T} does not necessarily imply~that
the associated bound state does not exist. Rather, it should be
understood as an indication that, within the adopted setting, no
trustable upper limit (apart from zero) could be found.

\begin{table}[ht]\caption{Variational upper bounds, in units of
the identical mass, $m$, of the two bound-state constituents, on
the binding energies (\ref{B}) of the low-lying bound states of
generic relativistic Hellmann problems (identifying any bound
state by radial quantum number $n_r$ and orbital angular momentum
quantum number $\ell$), derived by relying on the trial-space
basis (\ref{L}) and employing, as numerical input values, for the
two variational parameters $\beta=1$ and $\mu=m$, for the Yukawa
range parameter $b=m$, and three illustrative combinations of the
coupling strengths: $\kappa=\ups=\frac{1}{2}$, exemplifying all
set-(\ref{y=k}) potentials; $\kappa=1$ and~$\ups=-1$,~exemplifying
all set-(\ref{ye-k}) potentials; $\kappa=1$ and $\ups=-2$,
exemplifying all set-(\ref{yl-k}) potentials. (To facilitate
appreciation, the spectral lower bounds arising from
Eqs.~(\ref{SM}), (\ref{Se}), or (\ref{Sf}) are also provided.)}
\label{T}\begin{center}\begin{tabular}{cclll}\toprule
\multicolumn{2}{c}{Bound state}&
\multicolumn{1}{c}{$\quad\kappa=\ups=\frac{1}{2}\quad$}&
\multicolumn{1}{c}{$\kappa=1$, $\ups=-1$}&
\multicolumn{1}{c}{$\kappa=1$, $\ups=-2$}\\\cline{1-2}\\[-1.75ex]
$n_r$&$\ell$&\multicolumn{1}{c}{[case (\ref{y=k})]}
&\multicolumn{1}{c}{[case (\ref{ye-k})]}&\multicolumn{1}{c}{[case
(\ref{yl-k})]}\\\midrule\multicolumn{2}{l}{Spectral lower bound}&
$\quad$[Eq.~(\ref{SM})]& $\quad$[Eq.~(\ref{Se})]&
$\quad$[Eq.~(\ref{Sf})]\\[.5ex]\multicolumn{2}{r}{$\sigma(H)\;[m]$}
&$\quad-0.58578\dots$&$\quad-1$&$\quad-0.37336\dots$\\\midrule
0&0&$\quad-0.11673$&$\quad-0.17951$&$\quad-0.14410$\\
0&1&$\quad-0.01579$&$\quad-0.06294$&$\quad-0.06157$\\
0&2&$\quad-0.00616$&$\quad-0.02813$&$\quad-0.02812$\\
0&3&$\qquad$~~---&$\quad-0.01553$&$\quad-0.01553$\\[.5ex]
1&0&$\quad-0.02107$&$\quad-0.05464$&$\quad-0.04786$\\
1&1&$\quad-0.00509$&$\quad-0.02810$&$\quad-0.02762$\\
1&2&$\qquad$~~---&$\quad-0.01482$&$\quad-0.01481$\\
1&3&$\qquad$~~---&$\quad-0.00624$&$\quad-0.00624$\\[.5ex]
2&0&$\quad-0.00688$&$\quad-0.02566$&$\quad-0.02338$\\
2&1&$\qquad$~~---&$\quad-0.01391$&$\quad-0.01356$\\
2&2&$\qquad$~~---&$\quad-0.00122$&$\quad-0.00120$\\[.5ex]
3&0&$\qquad$~~---&$\quad-0.01104$&$\quad-0.00840$\\\bottomrule
\end{tabular}\end{center}\end{table}

\section{Exploiting Hellmann Potentials: Rules of the Game}
\label{I}For the notationally simpler case of equal-mass
constituents, we sketched the boundaries to be expected for
bound-state spectra of spinless relativistic Hellmann problems;
with a little more effort, the generalization to unequal masses is
straightforward. These findings~should, for future Hellmann
studies, reduce the risk to get lost in the vast jungle of
approximations.

Since recently, interest in the solutions of the spinless
\emph{relativistic\/} Hellmann problem~can be spotted in the
literature. At present, we are aware of just two attempts
\cite{Arda,Antia} to exploit for that purpose a longstanding but
very approximate way of dealing with spinless Salpeter equations.
Unfortunately, from our point of view this latter approach is very
problematic in (at least) two well-known respects: On the one
hand, as a first step towards treatability, the procedure relies
on an expansion of the square-root operator of any relativistic
kinetic term $T(\bm{p})$ in powers of $\bm{p}^2/m^2$ and
subsequent truncation of the expansion at a convenient order. The
\emph{pseudo}-spinless-Salpeter Hamiltonian emerging from such
mistreatment, however, can be shown \cite{WL:WS} to be an operator
unbounded from below for all interaction potentials that~are not
too singular at the spatial origin, including all generalized
Hellmann potentials. On the other hand, as a second step, the
formulation of an equation of Schr\"odinger type is enforced by
reinsertion of that eigenvalue equation of the
pseudo-spinless-Salpeter Hamiltonian that holds for its
lower-order truncation. The latter manipulation too is known to be
flawed,~due to an evidently inappropriate neglect of a commutator
term \cite{Chargui} in its derivation. In view~of these
deficiencies, it does not make sense to discuss the results of
Refs.~\cite{Arda,Antia} in any detail.


\small\begin{thebibliography}{30}
\bibitem{BSEa}H.~A.~Bethe and E.~E.~Salpeter, Phys.~Rev.~{\bf 82}
(1951) 309.
\bibitem{BSEb}M.~Gell-Mann and F.~Low,~Phys.~Rev.~{\bf 84} (1951)
350.
\bibitem{BSEc}E.~E.~Salpeter and H.~A.~Bethe, Phys.~Rev.~{\bf 84}
(1951) 1232.
\bibitem{WL:RVT}W.~Lucha and F.~F.~Sch\"oberl,
Phys.~Rev.~Lett.~{\bf 64} (1990) 2733.
\bibitem{WL:RVTs}W.~Lucha, Mod.~Phys.~Lett.~A {\bf 5} (1990) 2473.
\bibitem{WL:WS}W.~Lucha and F.~F.~Sch\"oberl, Int.~J.~Mod.~Phys.~A
{\bf 29} (2014) 1450057, arXiv:1401.5970~[hep-ph].
\bibitem{WL:Q@W}W.~Lucha and F.~F.~Sch\"oberl, EPJ Web Conf.~{\bf
80} (2014) 00049, arXiv:1407.4624 [hep-ph].
\bibitem{WL:H}W.~Lucha and F.~F.~Sch\"oberl, Int.~J.~Mod.~Phys.~A
{\bf 29} (2014) 1450181, arXiv:1408.4957~[hep-ph].
\bibitem{WL:Y}W.~Lucha and F.~F.~Sch\"oberl, Int.~J.~Mod.~Phys.~A
{\bf 29} (2014) 1450195, arXiv:1410.5241~[hep-ph].
\bibitem{WL:K}W.~Lucha and F.~F.~Sch\"oberl, Int.~J.~Mod.~Phys.~A
{\bf 30} (2015) 1550062, arXiv:1412.4950~[hep-ph].
\bibitem{H}H.~Hellmann, J.~Chem.~Phys.~{\bf 3} (1935) 61.
\bibitem{H+}H.~Hellmann and W.~Kassatotschkin, J.~Chem.~Phys.~{\bf
4} (1936) 324.
\bibitem{IWH}I.~W.~Herbst, Commun.~Math.~Phys.~{\bf 53} (1977) 285.
\bibitem{IWH+}I.~W.~Herbst, Commun.~Math.~Phys.~{\bf 55} (1977) 316.
\bibitem{MR}A.~Martin and S.~M.~Roy, Phys.~Lett.~B {\bf 233} (1989)
407.
\bibitem{RRSMS}J.~C.~Raynal, S.~M.~Roy, V.~Singh, A.~Martin, and
J.~Stubbe, Phys.~Lett.~B {\bf 320} (1994)~105.
\bibitem{WL:C}W.~Lucha and F.~F.~Sch\"oberl, J.~Math.~Phys.~{\bf
41} (2000) 1778, arXiv:hep-ph/9905556.
\bibitem{ICD}I.~Daubechies, Commun.~Math.~Phys.~{\bf 90} (1983)
511.
\bibitem{WL:RCP}W.~Lucha and F.~F.~Sch\"oberl, Phys.~Rev.~A {\bf
54} (1996) 3790, arXiv:hep-ph/9603429.
\bibitem{WL:T1}W.~Lucha and F.~F.~Sch\"oberl, Int.~J.~Mod.~Phys.~A
{\bf 14} (1999) 2309, arXiv:hep-ph/9812368.
\bibitem{WL:T2}W.~Lucha and F.~F.~Sch\"oberl, Fizika B {\bf 8}
(1999) 193, arXiv:hep-ph/9812526.
\bibitem{WL99}W.~Lucha and F.~F.~Sch\"oberl, J.~Math.~Phys.~{\bf
41} (2000) 1778, arXiv:hep-ph/9905556.
\bibitem{WL:TWR}W.~Lucha and F.~F.~Sch\"oberl, Recent
Res.~Dev.~Phys.~{\bf 5} (2004) 1423, arXiv:hep-ph/0408184.
\bibitem{HK}R.~L.~Hall and Q.~D.~Katatbeh, Phys.~Lett.~A {\bf 287}
(2001) 183, arXiv:math-ph/0107015.
\bibitem{VB}V.~Bargmann, Proc.~Natl.~Acad.~Sci.~USA {\bf 38} (1952)
961.
\bibitem{RS}M.~Reed and B.~Simon, \emph{Methods of Modern
Mathematical Physics IV: Analysis of Operators\/} (Academic Press,
New York, 1978).
\bibitem{AS}M.~Abramowitz and I.~A.~Stegun (eds.), \emph{Handbook
of Mathematical Functions\/} (Dover, New York, 1964).
\bibitem{Antia}A.~D.~Antia and E.~E.~Ituen,
Ann.~Univ.~Mar.~Curie-Sk\l odowska~(Lublin) {\bf 71}(AAA) (2016)
53.
\bibitem{Arda}A.~Arda, Indian J.~Phys.~{\bf 91} (2017) 903;
arXiv:1701.01336 [physics.gen-ph].
\bibitem{Chargui}Y.~Chargui, Eur.~Phys.~J.~Plus {\bf 133} (2018)
543.\end{thebibliography}
\end{document}